\documentclass[%
 reprint,
 amsmath,amssymb,
 aps,
]{revtex4-1}

\usepackage{graphicx}% Include figure files
\usepackage{dcolumn}% Align table columns on decimal point
\usepackage{bm}% bold math
\usepackage[utf8]{inputenc}
\usepackage[T1]{fontenc}
\usepackage{here}
\usepackage{units}

\begin{document}

%\preprint{APS/123-QED}

\title{Harmonic concatenation of 1.5-femtosecond-pulses in the deep ultraviolet}% Force line breaks with \\

\author{Jan Reisl\"ohner}
\author{Christoph Leithold}
\author{Adrian N. Pfeiffer}

\affiliation{Institute of Optics and Quantum Electronics, Abbe Center of Photonics, Friedrich Schiller University, Max-Wien-Platz 1, 07743 Jena, Germany}

%\affiliation[*]{Corresponding authors: jan.reisloehner@uni-jena.de, a.n.pfeiffer@uni-jena.de}

\date{\today}% It is always \today, today,

\begin{abstract}
Laser pulses with a duration of one femtosecond or shorter can be generated both in the IR-VIS and in the extreme UV, but the deep UV is a spectral region where such extremely short pulses have not yet been demonstrated.
Here, a method for the synthesis of ultrashort pulses in the deep UV is demonstrated, which utilizes the temporal and spatial harmonics that are generated by two noncollinear IR-VIS pulses in a thin MgF$_2$ plate. 
By controlling the groove-envelope phase of the IR-VIS pulses, spatial harmonics are concatenated to form deep UV waveforms with a duration of 1.5\,fs. 
\end{abstract}

\maketitle

%\tableofcontents

%%%%%%%%%%%%%%%%%%%%%%%%%%%%%%%%%%%%%%%%%%%%%%%%%%%%%%%%%%%%%%%%%%%%%%%%%%%%%%%%%%%%%%%%%%%%%%%%%%%%%

%%%%% Introduction
Waveforms with durations short enough to probe the electronic timescale can be generated both in the extreme UV and in the IR-VIS region of the electromagnetic spectrum. High-order harmonic generation can be used to generate isolated attosecond pulses and pulse trains in the extreme UV \cite{RN249}, and since very recently the coherent synthesis of optical pulses is used to generate subcycle optical waveforms in the IR-VIS region \cite{RN114}. In the deep UV (DUV), where many basic aromatic molecules absorb radiation and undergo photochemical reactions \cite{RN222}, pulses with such extremely short durations have not yet been demonstrated \cite{RN221}. Exploiting nondegenerate frequency mixing in a hollow waveguide \cite{RN261}, in filamentation \cite{RN265} or in a thin transparent solid \cite{RN263}, it has been possible to generate pulses with a duration on the order of 10\,fs \cite{RN261}. A similar pulse duration has been achieved by achromatic frequency doubling of visible ultrashort pulses \cite{RN264}. Also self-compression of DUV pulses by filamentation has been demonstrated, yielding comparatively short pulses of $\sim15\,$fs \cite{RN213}. The shortest pulse duration reported to date of 2.8\,fs has been achieved by frequency conversion in a gas cell \cite{RN212}, which, however, requires extremely short (<\,4\,fs) fundamental pulses in the IR-VIS. 

Despite these engagements of several groups in the development of pulse generation methods, spectroscopy with sub-10-fs DUV pulses has not often been reported to date. A fundamental difficulty is rooted in the large group velocity dispersion of any optical component in the DUV, causing a DUV pulse to be readily broadened and distorted before it reaches the sample. As a rare exception, it was recently demonstrated that $\sim\,$10-fs-DUV pulses, generated by chirped-pulse four-wave mixing \cite{RN262}, can be used in pump-probe spectroscopy of an aqueous solution of thymine \cite{RN222}. 

For applications like transient absorption spectroscopy, DUV pulses with durations shorter than currently available would be very useful. For example, in order to observe the transient change of band structures in dielectrics and the related screenings in the presence of a strong field pulse, probe pulses are required that i) have a duration shorter than the optical period of the strong field pulse, ii) encompass the wavelength of the dielectric bandgap, and iii) are available at the sample position. 

%%%%% Short summary
Here, a scheme for the coherent synthesis of DUV waveforms is presented. The principle is the concatenation of spatial and temporal harmonics that are created by two noncollinear IR-VIS pulses. Cascaded processes of frequency conversion and self-diffraction \cite{RN220} in a thin MgF$_2$ plate yield a multifaceted emission pattern of temporal and spatial harmonics. The emission angle of the harmonic orders depends on the frequency. This spatiotemporal coupling is exploited to concatenate two neighboring spatial harmonics by adjusting the crossing angle of the fundamental pulses. Broadband waveforms arise, of which the lower frequency bands belong to one spatial harmonic, and the higher frequency bands belong to the other spatial harmonic. In order to synthesize ultrashort waveforms from these frequency bands, their phases must be adjusted. This is accomplished by controlling the groove-envelope phase (GEP) \cite{RN94} of the fundamental pulses, which is determined by their delay on the subcycle timescale. Using temporal harmonics of third order, the concatenation of two spatial harmonics synthesizes DUV pulses of 1.5\,fs. A generic feature is that these waveforms are spatially separated from the fundamental pulses and are therefore available for immediate spectroscopic usage without any subsequent optical elements.

\begin{figure}[htbp]
\centering
\includegraphics[width=\linewidth]{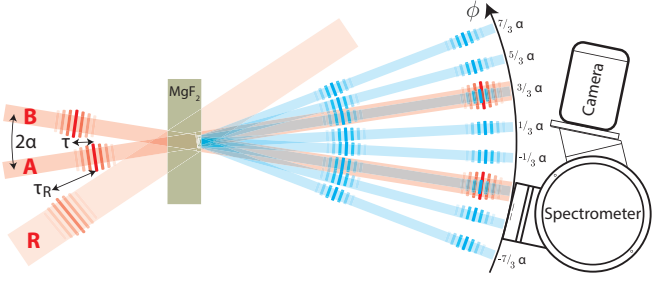}
\caption{Experimental setup. Pulses A and B generate DUV double pulses. Pulse R is overlapped only for the pulse characterization measurements.}
\label{fig:Setup}
\end{figure}

%%%%% Experimental Setup
Experimentally, two few-cycle VIS-IR pulses, labeled A and B (center wavelength $\lambda_{A,B}$ = 700\,nm, pulse duration $t^{FWHM}_{A,B}$ = 4.8\,fs, intensity $I_{A,B}$ = 4\,TW/cm$^2$), are focused noncollinearly with polarization perpendicular to the plane of incidence into a 100-$\mu$m-thick polycrystalline MgF$_2$ plate (half crossing angle $\alpha = 0.75^{\circ}$, beam waist $\sim$\,100\,$\mu$m) with variable delay $\tau$, see Fig.\,\ref{fig:Setup}. MgF$_2$ is chosen because of its wide bandgap (11.3\,eV) and the rather weak group velocity dispersion in the DUV. The experiment is carried out in vacuum, inhibiting nonlinear interactions with air. A spectrometer has been constructed with both spectral resolution and resolution in the emission angle $\phi$. 

\begin{figure}[htbp]
\centering
\includegraphics[width=\linewidth]{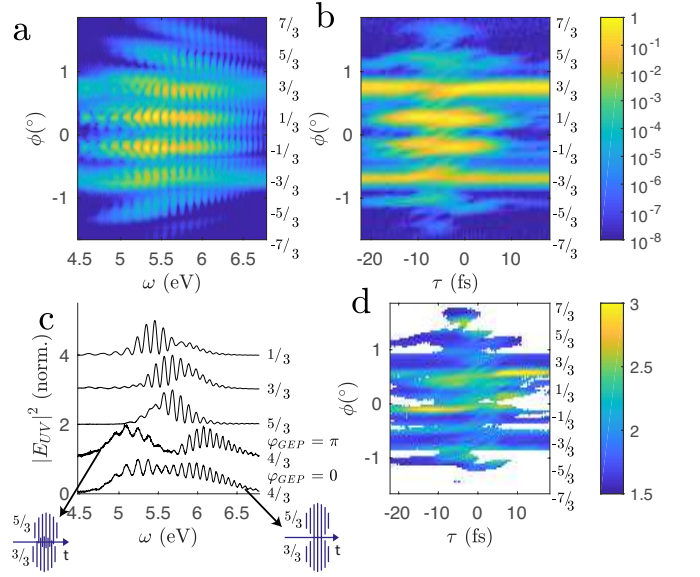}
\caption{Measured DUV intensities as function of $\phi$ and $\omega$ for $\tau$\,=\,0 (a) and as function of $\phi$ and $\tau$ at $\omega$\,=\,5.4\,eV (b). Spectra at $\phi$\,=\,\nicefrac{1}{3}\,$\alpha$,\,\nicefrac{3}{3}\,$\alpha$,\,\nicefrac{5}{3}\,$\alpha$ are shown for $\varphi_{GEP}$\,=\,0 and at $\phi$\,=\,\nicefrac{4}{3}\,$\alpha$ for $\varphi_{GEP}$\,=\,0 and $\pi$ (c). The Fourier limit of the pulse duration is displayed in (d) in areas where the normalized intensity is greater 10$^{-4}$.}
\label{fig:Data}
\end{figure}

%%%%% Experimental Data
A multifaceted emission pattern of temporal and spatial harmonics is generated through cascaded processes of frequency conversion and self-diffraction. At DUV frequencies, which corresponds to temporal harmonics of third order, several orders of self-diffraction (spatial harmonics) can be identified, see Fig.\,\ref{fig:Data}. The spatial harmonics with emission angle equal to one of the two fundamental pulses ($\phi$\,=\,$\pm$\,\nicefrac{3}{3}\,$\alpha$) is observed at all pulse delays. Within the bisector of the two generating beams ($\phi$\,=\,$\pm$\,\nicefrac{1}{3}\,$\alpha$), spatial harmonics generation is restricted to temporal pulse overlap. Outside the bisector ($\phi$\,=\,$\pm$\,\nicefrac{5}{3}\,$\alpha$ and $\phi$\,=\,$\pm$\,\nicefrac{7}{3}\,$\alpha$), spatial harmonics generation is efficient only within a region of very small pulse delays. In between the spatial harmonics, interferences arise that are dependent on the pulse delay $\tau$ on the subcycle timescale.

%%%%% Maker Fringes/Double pulse
With the restriction to 1D pulse propagation in $z$-direction, the DUV field $E_{\mathtt{UV}}$ is given by (atomic units are used and the convention for Fourier transform is $\mathcal{F} \{ f(t) \} \propto \int_{-\infty}^{+\infty} f(t) \mathrm{e}^{-i \omega t} \mathrm{d}t $):
\begin{equation} 
E_{\mathtt{UV}}(\omega,z) = - i \frac{2\pi\omega}{cn(\omega)} \mathrm{e}^{-i n(\omega) \frac{\omega}{c} z} \int_{0}^{z} dz' P(\omega,z') \mathrm{e}^{i n(\omega) \frac{\omega}{c} z'}),
\label{eq:PulseProp1D}
\end{equation}
where $\omega$ is the angular frequency (resp. photon energy), $c$ is the speed of light, $n(\omega)$ is the refractive index and $P$ is the nonlinear polarization response. Within the approximation that the fundamental pulses propagate linearly and that the refractive index is constant at fundamental frequencies ($n(\omega_{IR}) = n_{IR}$), Eq.\,(\ref{eq:PulseProp1D}) is solved by:
\begin{equation} 
E_{\mathtt{UV}}(\omega,z) = - \frac{2 \pi P(\omega,z=0)}{n(\omega) \left( n(\omega)\!-\!n_{IR} \right)}  
\left( \mathrm{e}^{-i n_{IR} \frac{\omega}{c} z} - \mathrm{e}^{-i n(\omega) \frac{\omega}{c} z} \right). 
\label{eq:Solve1D}
\end{equation}
Fringes appear in the spectrum $ \left| E_{\mathtt{UV}}\right|^2 $, which are observed in the data at all spatial harmonics (Fig. \ref{fig:Data}). This phenomenon is usually avoided by phase-matching methods to increase the DUV pulse energy, but here the focus is on shortening the pulse duration. In time domain, the spectral fringes correspond to a DUV double pulse. This might be surprising, because one might expect that a single broadened pulse is generated instead of two well-separated pulses, but the pulse splitting has been described before \cite{RN231}, and it directly follows from Eq.\,(\ref{eq:Solve1D}). At the end of the MgF$_2$ plate, the leading and the trailing DUV pulses are well separated by $\sim$\,36\,fs. Remarkably, only the trailing pulse (represented by $\mathrm{e}^{-i n(\omega) \frac{\omega}{c} z}$) is broadened by the strong group velocity dispersion in the DUV. The leading pulse (represented by $\mathrm{e}^{-i n_{IR} \frac{\omega}{c} z}$) propagates at the speed of the IR-VIS pulse and remains short throughout the propagation with a pulse duration of $t^{FWHM}_{A,B}$ /$\sqrt{3}$ = 2.8\,fs in the case of instantaneous third-order polarization response.

%%%%% Groove-envelope phase
Let $E_\mathtt{A}(t,x)= \Re \{ f_\mathtt{A}(t,x)\exp(i(\omega_0 t - k_x x + \varphi_\mathtt{CEP} )) \}$ and $E_\mathtt{B}(t,x)= \Re \{ f_\mathtt{B}(t-\tau,x)\exp(i(\omega_0 (t-\tau) + k_x x + \varphi_\mathtt{CEP} )) \}$ be the electric fields of laser pulses A and B respectively inside the nonlinear medium at $z=0$. The carrier-envelope phase (CEP) $\varphi_\mathtt{CEP}$ determines the temporal position of the carrier wave underneath the temporal envelope. The CEP is not stabilized from shot-to-shot, but is identical for A and B in each shot. At temporal pulse overlap, an intensity grating (laser induced grating) appears with groove spacing $\pi / k_x \, \approx$\,20\,$\mu$m. The GEP $\varphi_\mathtt{GEP} = \omega_0 \tau$ is the phase between the grooves of the intensity grating and the spatial envelope of the beams \cite{RN94}. The GEP is adjusted by the pulse delay $\tau$, where a delay of one optical cycle translates into a GEP shift of 2$\pi$. The GEP is the spatial analogue to the CEP: The CEP determines the interference of converted temporal frequencies (multiples of $\omega_0$), while the GEP determines the interference of converted spatial frequencies (multiples of $k_x$) \cite{RN94}. In order to observe CEP dependence, the spectral width must be large enough for spectral overlap of neighbouring temporal harmonics (for example in an $f$-2$f$ interferometer, where a pulse with an octave-spanning spectrum interferes with a frequency-doubled replica). In order to observe GEP dependence, the divergence of the spatial harmonics must be large enough for overlap of neighbouring spatial harmonics. This is adjusted by the crossing angle and beam waist.

For close-to-collinear configurations, the intensity grating consists of only a few grooves, and phenomena of nonlinear optics yield subcycle-dependent (GEP-dependent) signals. Recently it has been demonstrated that a close-to-collinear geometry can be exploited to achieve subcycle resolution in probe retardation measurements \cite{RN66} and self-diffraction \cite{RN94}. Self-diffraction denotes that A and B (or waves that are generated by A and B) are diffracted on the laser induced grating. A detailed discussion of the timing of the diffraction orders was presented in Ref.\,\cite{RN94}. If the GEP is shifted by $\delta$, then the phase of the DUV light (temporal harmonics of third order) in spatial harmonic $\phi$\,=\,\nicefrac{1}{3}\,$\alpha$ is shifted by $\delta$, in spatial harmonic $\phi$\,=\,-\nicefrac{1}{3}\,$\alpha$ by $2 \delta$, and in spatial harmonic $\phi$\,=\,-\nicefrac{3}{3}\,$\alpha$ by $3 \delta$. This scheme continues outside the bisector. 

%%%%% Harmonic concatenation
Harmonic concatenation exploits that the harmonic orders exhibit spatiotemporal couplings: the emission angle depends on the frequency. For light in between two spatial harmonics, for example at $\phi$\,=\,\nicefrac{4}{3}\,$\alpha$, the lower frequency band belongs to harmonic $\phi$\,=\,\nicefrac{3}{3}\,$\alpha$, whereas the higher frequency band belongs to harmonic $\phi$\,=\,\nicefrac{5}{3}\,$\alpha$, see Fig.\,\ref{fig:Data}. The spectral bandwidth at $\phi$\,=\,\nicefrac{4}{3}\,$\alpha$ is much broader than at the spatial harmonics. Whereas the spectra of the spatial harmonics depend only very weakly on the GEP, the GEP dependence is very pronounced in between the orders. At $\phi$\,=\,\nicefrac{4}{3}\,$\alpha$, the spectral content at the center frequency can be tuned from constructive interference at $\varphi_\mathtt{GEP} = 0$ to destructive interference at $\varphi_\mathtt{GEP} = \pi$. In order to exploit the broad bandwidth for the synthesis of a short waveform, the frequency bands of the spatial harmonics must be concatenated by adjusting the crossing angle and the GEP. A first glimpse on achievable pulse durations is provided by the Fourier limit (Fourier transforms of the spectra with the assumption of a flat phase) after removal of the spectral fringes, see Fig.\,\ref{fig:Data}(d). The broadest bandwidths and potentially shortest pulses are located in between the spatial harmonics. 

\begin{figure}[htbp]
\centering
\includegraphics[width=\linewidth]{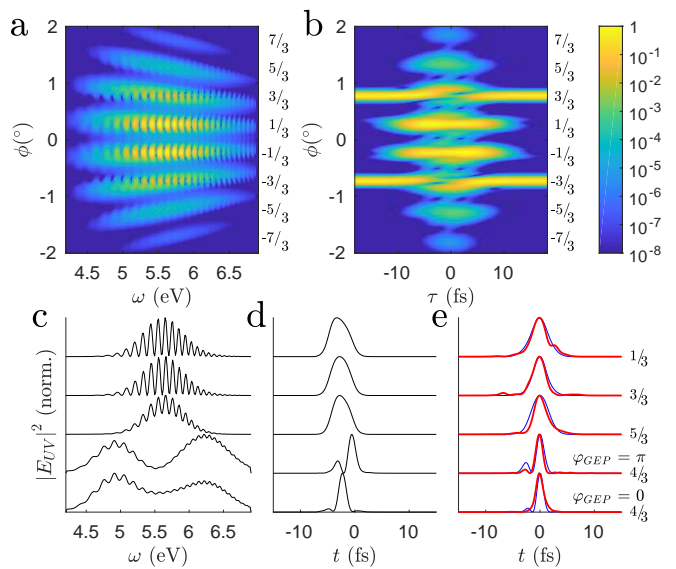}
\caption{Panels (a)-(c) are extracted from the simulation and correspond to the respective panels in Fig.\,\ref{fig:Data}. For the spectra shown in (c), the intensity envelopes of the leading pulse are depicted in (d). The result of the pulse characterization is shown in (e) for the measured data (red) and for the simulation (blue). The values for $\phi$ and $\tau$ in (c)-(e) are indicated on the right in (e).}
\label{fig:Simulation}
\end{figure}

%%%%% Simulation
In order to scrutinize the mechanism of harmonic concatenation, simulations are employed similar as described in Refs.\,\cite{RN94, RN266, RN117, RN258}. The unidirectional pulse propagation equation (UPPE) is integrated numerically using the split-step method. One transverse dimension (the $x$-dimension) is included to account for the noncollinear geometry. The electric field is treated as scalar field, because all pulses are polarized perpendicular to the plane of incidence. To initialize the fundamental fields A and B at the beginning of the MgF$_2$ plate, the pulse retrieval of the data (described in the next paragraph) is employed: for all fundamental fields, the cube root of the complex envelope of the leading pulse in spatial harmonic $\phi$\,=\,\nicefrac{3}{3}\,$\alpha$ is used. The pulse duration of the fundamental pulses is 4.8\,fs, which was additionally confirmed with a third-order harmonic FROG\,\cite{RN108}. Numerical refractive index data is used for the linear response, and the nonlinear response is calculated assuming instantaneous response with $\chi^{(3)} = 0.82\ \mathrm{au}$, corresponding to a nonlinear refractive index n$_2$\,=\,0.057\,cm$^2$/PW \cite{RN267}. Subsequent to the sample propagation, linear propagation into the far field is performed. The UPPE simulations reproduce the experimental data very well, see Fig.\,\ref{fig:Simulation}(a)-(c). In addition to the spectral intensities, the time-domain representation of the fields can be extracted from the simulation. At all emission angles, well-separated double pulses are found, in accordance with the 1D propagation described by Eq.\,(\ref{eq:Solve1D}). At spatial harmonics $\phi$\,=\,\nicefrac{1}{3}\,$\alpha$,\,\nicefrac{3}{3}\,$\alpha$,\,\nicefrac{5}{3}\,$\alpha$, the pulses are independent on the GEP and the leading pulses have a pulse duration of $\sim$\,3\,fs. Shorter pulses are found in between the spatial harmonics, but their shape depends on the GEP. At $\phi$\,=\,\nicefrac{4}{3}\,$\alpha$, the case of destructive interference ($\varphi_\mathtt{GEP} = \pi$, respectively $\tau$\,=\,1.3\,fs) leads to a pulse that is distorted. The reason is that the frequency bands of the spatial harmonics are not concatenated with a GEP that is ideal for the synthesis of a short waveform. Better suited for ultrafast spectroscopy is the pulse that is generated in the case of constructive interference ($\varphi_\mathtt{GEP} = 0$, respectively $\tau$\,=\,0). Here, the pulse is only very weakly distorted, is well-separated from the trailing pulse and has a duration of only 1.5\,fs. 

\begin{figure}[htbp]
\centering
\includegraphics[width=\linewidth]{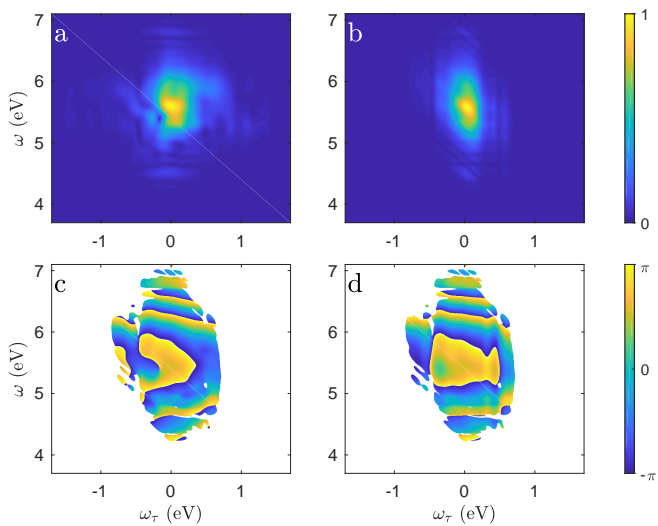}
\caption{Magnitudes (a), (b) and phases (c), (d) of experimentally measured (a), (c) and retrieved (b), (d) traces $\textbf{Y}_{\mathrm{c}}(\omega,\omega_\mathtt{R})$ used for the pulse characterization at $\phi$\,=\,\nicefrac{4}{3}\,$\alpha$ and $\tau$ = 0 (depicted in Fig.\,\ref{fig:Simulation}(e)). Phase values are only shown where the normalized magnitudes are greater than 0.01 in both the measured and retrieved traces.}
\label{fig:Traces}
\end{figure}

%%%%% Pulse characterization
For a further validation of the simulation results, the DUV waveforms are characterized with the recently developed method cross-phase modulation scans \cite{RN266}. Spectra $ \left| E_{\mathtt{UV}} \right|^2 $ are recorded, while a further VIS-IR pulse, labeled R, is overlapped at a variable delay $\tau_\mathtt{R}$, see Fig.\,\ref{fig:Setup}. The pulse characterization is performed both experimentally and on synthetic data, which is generated using the UPPE simulation where all three pulses A, B and R are initialized. The variant \emph{center} is used, where the phase of the leading DUV pulse is shifted via cross-phase modulation. As a preparation for the pulse retrieval, the spectra $ \left| E_{\mathtt{UV}} \right|^2 $ at each $\tau_\mathtt{R}$ are inverse Fourier transformed, the alternating component (side peak) is selected and shifted to zero frequency, and thereafter Fourier transformed. The basis for pulse retrieval is the data trace $\textbf{Y}_{\mathrm{c}}(\omega,\omega_\mathtt{R})$, which is calculated by subtraction of the non-$\tau_\mathtt{R}$-dependent background, followed by Fourier transform from $\tau_\mathtt{R}$ to $\omega_\mathtt{R}$. The pulse retrieval from $\textbf{Y}_{\mathrm{c}}(\omega,\omega_\mathtt{R})$ is analytic, and the fidelity of the retrieval is checked by comparing the complex-valued data trace with the retrieved trace (Fig.\,\ref{fig:Traces}). The retrieved pulses are depicted in Fig.\,\ref{fig:Simulation}(e) for both the experimental and the synthetic data. The pulse shapes predicted by the simulation (Fig.\,\ref{fig:Simulation}(d)) are reasonably well reproduced both applying the retrieval to the synthetic data (Fig.\,\ref{fig:Simulation}(e), blue) and to the experimental data (Fig.\,\ref{fig:Simulation}(e), red).

%%%%% Discussion
A distinct advantage of harmonic concatenation is that the generated waveforms are spatially separated from the fundamental pulses. The generated waveforms are not well suited to act as pump pulses, because they are very weak and the spatiotemporal coupling prevents refocusing. They are, however, excellently suited to act as probe pulses. A sample to be investigated can be placed directly after the generation medium, where a pump pulse can be overlapped. In this scheme, the probe pulses generated by harmonic concatenation are available for immediate spectroscopic usage without any subsequent optical elements, which have previously been the main obstacle for DUV spectroscopy in the sub-10-fs range \cite{RN222}. As the trailing pulse is well separated with a large inter-pulse delay, it may enable simultaneous transient absorption and dispersion measurements. If the trailing pulse should be suppressed, for example when a process with a decay time larger than the inter-pulse delay is studied, this could be done by deflecting it with the aid of a spatially shaped pulse. In the present setup, pulse R could be shaped to have an intensity gradient extending over the beam profiles of A and B and be temporally overlapped with the trailing pulse. Cross-phase modulation would induce a spatial gradient in the phase of the trailing pulse, effectively deflecting the trailing pulse.

%%%%% Conclusion
In conclusion, harmonic concatenation is a method for the synthesis of short waveforms in the DUV. It requires two noncollinear IR-VIS pulses that trigger cascaded processes of frequency conversion and self-diffraction in a thin dielectric. The spatiotemporal coupling of the temporal and spatial harmonics is exploited to concatenate two neighboring spatial harmonics. This requires adjusting the crossing angle of the fundamental pulses and their GEP. Using temporal harmonics of third order generated in a 100-$\mu$m-MgF$_2$ plate by 4.8\,fs-pulses, the concatenation of two spatial harmonics synthesizes DUV pulses of 1.5\,fs.

%%%%%%%%%%%%%%%%%%%%%%%%%%%%%%%%%%%%%%%%%%%%%%%%%%%%%%%%%%%%%%%%%%%%%%%%%%%%%%%%%%%%%%%%%%%%%%%%%%%%%

\section*{Funding.}
Deutsche Forschungsgemeinschaft (DFG) (PF 887/1-1); Daimler und Benz Stiftung (32 07/15); Europäische Fonds für regionale Entwicklung (EFRE) Thüringen (2016 FGI 0023); Carl-Zeiss-Stiftung.

%\section*{Acknowledgment.} 
%A. Pfeiffer acknowledges support from the Carl Zeiss Foundation.

%%%%% Bibliography
\bibliography{exportlist}
%\bibliographyfullrefs{exportlist}

\end{document}